\documentclass[doublecol]{epl2} 
\usepackage{graphicx}
\usepackage{dcolumn}
\usepackage{bm}
\usepackage{amssymb}
\usepackage{amsmath}
\begin{document}
\setcounter{page}{1}
\vskip 2cm
\title{Black-Hole evaporation from the perspective of neural networks}
\shorttitle{\textbf{Black-Hole evaporation from the perspective of neural networks}}

\author{Ivan Arraut}
\shortauthor{Ivan Arraut}
\institute{The Open University of Hong Kong, 30 Good Shepherd Street, Homantin, Kowloon}
 
\pacs{04.62.+v}{Quantum fields in curved spacetime}
\pacs{04.70.Dy}{Quantum aspects of black holes, evaporation, thermodynamics}
\pacs{03.75.Nt}{Other Bose-Einstein condensation phenomena}

\abstract{We study the black-hole evaporation from the perspective of neural networks. We then analyze the evolution of the Hamiltonian, finding in this way the conditions under which the synapse connecting the neurons changes from gravitatory to inhibitory during the evaporation process.}
 
\maketitle
 
\maketitle 
\section{Introduction}	
From the classical point of view, the Black-Holes are objects able to absorb particles (information) which can never escape away \cite{Clas}. In addition, they can store information in the most efficient way \cite{Dvali}. If we make an analogy with the way how a system of neural networks administrates the information, this means a perfect storage \cite{Dvali}. The analogy between Black-Holes and neural networks, was explored in \cite{Dvali}. In such a case it was also suggested that the black-holes can be visualized as a critical state of a group of attractive bosons. The critical condition corresponds to the situations where the phenomena of assisted gaplessness occurs and helps to cancel out the gap of the corresponding assisted modes. This happens when a highly occupied master mode helps other modes to become gapless. A neural network experiencing assisted gaplessness can store a huge amount of patterns at a costless price when the critical condition is reached. This effect is related to the phenomena of classicalization introduced in \cite{DV2}. Interesting aspects about the physics of Black-Holes as well as the possible solution to some important physical problems, like the UV-completion of gravity are also explored inside this formulation \cite{UVcom}. Exploring the UV-completion of gravity via classicalization is different to the traditional way of approaching to this problem \cite{Trad}. Through this formulation, an alternative solution to the hierarchy problem has been proposed \cite{Dvali, Hyer}. In this paper we explore the Black-Hole evaporation from the perspective of neural networks. When the Black-Hole evaporates, it loses information at a rate defined by the Hawking radiation \cite{Hawking}. Here we demonstrate that when there is a simultaneous evaporation of the master and the assisted modes, the synapse becomes less gravitatory in time and eventually, under some special conditions, it might become inhibitory. We obtain the conditions under which the synapse becomes inhibitory during the evaporation process.          

\section{Black-Holes as a critical state of a Bosonic field}   \label{BHBF}

We can consider a Bosonic field obeying the usual commutation relations defined as

\begin{equation}
[\hat{a}_i, \hat{a}^+_j]=\delta_{ij},\;\;\;\;\;[\hat{a}_j, \hat{a}_k]=[\hat{a}_j^+, \hat{a}_k^+]=0.
\end{equation}
The level of excitation of a mode in the state $\vert n_k>$ is characterized by the occupation number defined as $\hat{n}_k=\hat{a}^+_k\hat{a}_k=0, 1, ..., (d-1)$. There are $K+1$ possible degrees of freedom (oscillators), each one forming a qdit. The most important quantity to be analyzed is the energy gap between the states $\vert n_k>$ and $\vert n_k\pm 1>$, here defined as $E_{n_0, n_1, ..., n_{K}}=\sum_k E_kn_k$. This quantity marks the effort necessary for the system to store some information patterns. We consider $\vert n_k>$ as a qdit with $d-1$ different possible states \cite{Dvali}. Each quantum occupation number corresponds to a neuron which can be excited by some input stimuli. When the occupation number increases, then the gap associated to the neuron also increases correspondingly. If we have $K+1$ of such neurons (states), then the state representing the information storage pattern is given by the tensorial product of the particle numbers (gap) associated to each neuron and here simplified as $\vert n_0, n_1, n_2,..., n_{K}>$ \cite{Dvali}. The neuron excitation is defined by the level of occupation number which can reach until $d-1$.    

\subsection{The phenomena of assisted gaplessness}   \label{assisted}

The phenomena of assisted gaplessness can be understood if we take the master mode to be defined by the occupation number $\hat{n}_0$. Then the Hamiltonian is given by 

\begin{equation}   \label{Hamiltton}
\hat{H}=\sum_{k=1}^KE_k\left(1-\alpha\hat{n}_0\right)\hat{n}_k+E_0\hat{n}_0+...,
\end{equation}           
In this case, the master mode is taken to be highly occupied and the coupling with this mode and the others is "gravitatory" (negative coupling). As far as the coupling between the master modes and the others is negative, then the synapse connection is considered as excitatory or energetically favorable. Defining the occupation number of the master mode as $<\hat{n}_0>=N_0$, then the effective gap for the other modes is given by \cite{Dvali}

\begin{equation}   \label{gap}
E_k^{eff}=E_k\left(1-\lambda\right),
\end{equation}
with $\lambda=\alpha N_0$. Then the presence of a master mode, helps the other modes to decrease the gap such that the excitation becomes easier. The critical condition, which is the one corresponding to the black-hole case is given by the value of the parameter $\lambda=1$. In this case, the assisted modes become gapless and it costs zero energy to store patterns inside them. 
Note that the critical condition is equivalent to $N_0=\lambda^{-1}$. In such a case there is a degeneracy in all the states $\vert N_0, n_1, n_2,..., n_{K}>$ and given by $d^{K}$. This degeneracy is connected with the definition of entropy. This previous degeneracy for example explains why the entropy of the black-holes scale with the area law. 

\section{The Black-Hole evaporation: Standard case}   \label{BHBF2}

The Black-Hole evaporation process in its standard version connects two different definitions of vacuum by using the Bogoliubov transformations \cite{Hawking}. The original calculation developed by Hawking suggested a connection between the vacuum defined at the infinite future and another one defined in the past infinity. The idea in its simplest version is that if we have a scalar field expanded as

\begin{equation}   \label{Thia}
\phi=\sum_i\left(f_i\hat{a}_i+\bar{f}_i\hat{a}_i^+\right),
\end{equation} 
in the past null infinity which is devoid of particles, then we can connect the vacuum defined by $\hat{a}_i\vert 0>=0$ in the past null infinity with the vacuum defined in the future null infinity as $\hat{b}_i\vert\bar{0}>=0$. The massless scalar field $\phi$ can be defined everywhere in agreement with the expression (\ref{Thia}), or equivalently it can also be expanded as

\begin{equation}
\phi=\sum_i\left(p_i\hat{b}_i+\bar{p}_i\hat{b}_i^++q_i\hat{c}_i+\bar{q}_i\hat{c}_i^+\right),
\end{equation}  
with the modes $p_i$ defined in the future null infinity and the modes $q_i$ defined on the event horizon. Note that the information problem in a black-hole comes out precisely from the fact that only part of the information of the scalar field is contained inside the modes $p_i$, which are at the end the ones measured by the observer located in the asymptotic null infinity \cite{IP}. It is easy to demonstrate then that the vacuum defined by $\hat{a}_i\vert 0>=0$ and the one defined by $\hat{b}_i\vert\bar{0}>=0$ are connected by Bogoliubov transformations in order to preserve the canonical commutation relations. The particle creation effect then happens when the Bogoliubov coefficient mixing the positive and negative frequency modes does not vanish. Following the notation of \cite{Hawking}, we can write 

\begin{equation}    \label{Thia2}
\hat{b}_i=\sum_j\left(\bar{\alpha}_{ij}\hat{a}_j-\bar{\beta}_{ij}\hat{a}^+_j\right).
\end{equation}       
Here $\bar{\beta}_{ij}$ is the coefficient mixing positive and negative frequencies. Then an initial vacuum $\hat{a}_i\vert 0>=0$, is perceived in the future null infinity as a state full of particles, with the expectation value on the number of particles defined by 

\begin{equation}
<0\vert \hat{b}_i^+\hat{b}_i \vert0>=\sum_j\vert\beta_{ij}\vert^2.
\end{equation}
Then we conclude that an observer located in the future null infinity perceives the existence of particles coming from the black-hole. Here we will not derive explicitly the Bogoliubov coefficients, we are rather interested in the relation between them. In fact, the coefficient mixing positive and negative frequency modes $\beta_{ij}$, is related to the coefficient $\alpha_{ij}$ by the relation

\begin{equation}
\vert\alpha_{\omega\omega'}^{(2)}\vert=e^{\frac{\pi\omega}{\kappa}}\vert\beta_{\omega\omega'}^{(2)}\vert.
\end{equation}
Here the superindex $(2)$ means that we are considering the portion of the field which is able to enter the black-hole for being emitted again as radiation. From the Mathematical perspective the previous relation between the Bogoliubov coefficients creates the black-body radiation effect. Then whenever this relation is satisfied, we are recreating the effects of particle creation of Black-holes.  

\section{Black-Hole evaporation from the perspective of neural networks}   \label{BHBF4}

Knowing the relation between the Bogoliubov coefficients, necessary for obtaining the Black-Hole evaporation effects, we can then proceed to explore what are the consequences of the same relation inside a system experiencing assisted gaplessness. Here we use the analogy between a bosonic gas and a neural network proposed in \cite{Dvali}, in order to demonstrate the equivalent effect from this perspective. The analogy between a gas of Bosons and a neural network suggests the following dictionary \cite{Dvali}: 1). The neuronal excitation is equivalent to the particle occupation number of the bosonic field. 2). The particle momentum is associated to neural degree of freedom. 3). The oscillator energy gap is related to the excitation energy threshold of the neuron. 4). The gravitational interaction  is simulated by the excitatory synapse connection. 5). The Black-hole state is related to the Critical state of highly excited low threshold neurons. Considering this dictionary, we can conclude that the Hamiltonian (\ref{Hamiltton}) also represents the behavior of a neural network with one master neuron (mode). Here we want to analyze what happens if the different modes start to evaporate. In order to achieve this, we have to start analyzing the vacuum conditions, such that we are able to understand how the particles are detected in the future null infinity from the perspective of an observer located there. Although the evaporation process from the perspective of neural networks is interpreted in a different way, the same Mathematical tools are still valid in such scenario.

\subsection{Particles perceived in different vacuums}

In the language of the Black-Holes, the Hamiltonian (\ref{Hamiltton}) is expanded in the modes defined in the asymptotic past infinity, and then for the vacuum defined in such a case, we get 

\begin{equation}
\hat{H}\vert0>=0,
\end{equation}
which is a natural consequence of the vacuum condition $\hat{a}_i\vert0>=0$. On the other hand, the same Hamiltonian, could be expressed in terms of the modes defined at the future null infinity. In such a case, the vacuum is defined by $\hat{b}_i\vert\bar{0}>=0$. In addition, the Hamiltonian (\ref{Hamiltton}) expanded in terms of these modes, also satisfies the condition $\hat{\bar{H}}\vert\bar{0}>=0$. In asymptotically flat situations, the observers perceiving the black-hole evaporation, are usually located at the future null infinity. Taking into account this aspect, we then proceed to evaluate how an observer located at the asymptotic null infinity, would perceive the information stored in the Hamiltonian (\ref{Hamiltton}), but evaluated with respect to the vacuum where the information is originally stored, namely the vacuum obeying $\hat{a}_i\vert0>=0$. Since the observers located at the future null infinity define the modes in agreement with the operators $\hat{b}_i$ and $\hat{b}_i^+$, then we need to use the Bogoliubov transformation (\ref{Thia2}) inside (\ref{Hamiltton}) in order to describe the operators $\hat{b}_i$ in terms of the operators $\hat{a}_i$. Considering the vacuum expectation value $<0\vert\hat{\bar{H}}\vert0>$, some of the relevant terms to be evaluated are

\begin{equation}
\hat{b}_{0\omega}^+\hat{b}_{0\omega}=\int d\omega'\vert\beta^{(0)}_{\omega, \omega'}\vert^2,
\end{equation} 
where we have ignored all the terms which vanish after taking the vacuum expectation value. The previous result is the trivial one and it corresponds to the part of the Hamiltonian $E_0\hat{n}^{e}_0= E_{0}\int d\omega'\vert\beta_{\omega, \omega'}^{(0)}\vert^2$. We use the superindex $e$ because we want to remark that this quantity corresponds to the emission of particles from the Black-Hole. This part corresponds to the evaporation of the master modes without considering the coupling to the assisted modes. Analogously, the uncoupled part of the assisted modes evaporates in the same way if we consider $\hat{b}_{k\omega}^+\hat{b}_{k\omega}=\int d\omega'\vert\beta^{(k)}_{\omega, \omega'}\vert^2$, which helps us to define the term $E_{k}\hat{n}_{k}^{e}= E_{k}\int d\omega'\vert\beta^{(k)}_{\omega, \omega'}\vert^2$. Here again we use the superindex $e$. Finally, the interesting part comes from the assisted modes coupled to the master one. This part requires the evaluation of terms of the form

\begin{eqnarray}   \label{everla}
<0\vert\hat{b}^+_0\hat{b}_0\hat{b}^+_k\hat{b}_k\vert0>=\;\;\;\;\;\;\;\;\;\;\;\;\;\;\;\;\;\;\;\;\;\;\;\;\;\;\;\;\;\;\;\;\;\;\;\;\;\;\;\;\;\;\;\;\;\;\;\nonumber\\
<0\vert\left(\beta^{(0)}_{\omega, \omega'}\beta^{*(k)}_{\omega_k, \omega_k''}\alpha^{*(0)}_{\omega, \omega''}\alpha^{(k)}_{\omega_k, \omega_k'}\hat{a}_{0\omega'}\hat{a}_{0\omega''}\hat{a}^+_{\omega_k'}\hat{a}^+_{\omega_k''}\right)\vert0>\nonumber\\
+<0\vert\left(\vert\beta^{(0)}_{\omega, \omega'}\vert^2\vert\beta^{(k)}_{\omega_k, \omega_k'}\vert^2\hat{a}_{0\omega}\hat{a}^+_{0\omega'}\hat{a}_{k\omega_k}\hat{a}^+_{k\omega_k'}\right)\vert0>.
\end{eqnarray} 
Note that the first term of the right-hand side of this equation vanishes if the eigenstates generated by $\hat{a}_0^+\vert0>$ and $\hat{a}^+_k\vert0>$ are orthogonal. However, if there is some overlap between these states, then this term does not vanish in general. The second term in the sum never vanishes, independent on whether or not the master modes and the assisted ones are overlapped. The result (\ref{everla}) shows that the coefficient which does not mix the frequencies, namely, $\alpha_{\omega, \omega'}$, might also contribute to the emission of particles due to the coupling between the master modes and the assisted ones. This only happens when there is an overlap between the eigenstates related to the master modes with the eigenstates related to the assisted ones as has been just remarked.    

\subsection{Changes in the gap} 
 
The changes in the gap can be found from the net Hamiltonian, which is the original Hamiltonian with the subtractions coming from the terms related to the emission of particles just previously calculated. In fact, during the process of emission of particles, the Hamiltonian (\ref{Hamiltton}), becomes

\begin{eqnarray}   \label{Hal}
\hat{H}=\sum_{k=1}^K E_k\left(1-\alpha\left(\frac{\hat{n}_0\hat{n}_k-\hat{n}_0^e\hat{n}_k^e-\epsilon\hat{n}_{0, k\; mix}^{e\beta}\hat{n}^{e\alpha}_{0, k\; mix}}{\hat{n}_k-\hat{n}_k^e}\right)\right)\times\nonumber\\
(\hat{n}_k-\hat{n}_k^e)+E_0\left(\hat{n}_0-\hat{n}_0^e\right)+...
\end{eqnarray} 
Here we have defined 

\begin{eqnarray}   \label{security}
\hat{n}_0^e=\int d\omega' \left\vert\beta^{(0)}_{\omega, \omega'}\right\vert^2,\;\;\;\;\;\;\;\;\;\;\;\;\;\;\;\;\;\;\;\;\; \hat{n}_k^e=\int d\omega' \left\vert\beta^{(k)}_{\omega, \omega'}\right\vert^2,\;\;\nonumber\\
\hat{n}_{0, kmix}^{e\beta}=\int d\omega_{m}\beta^{(0)}_{\omega, \omega'}\beta^{*(k)}_{\omega_k, \omega_k''},\;\; \hat{n}^{e\alpha}_{0, kmix}=\int d\omega_{m}\alpha^{*(0)}_{\omega, \omega''}\alpha^{(k)}_{\omega_k, \omega_k'}.
\end{eqnarray}
In addition, in order to simplify the notation in the equations, we write the inverse operator $(\hat{n}_k-\hat{n}_k^e)^{-1}=\frac{1}{\hat{n}_k-\hat{n}_k^e}$, following the notation of ordinary numbers. The dimensionless parameter $\epsilon<<1$ introduced in eq. (\ref{Hal}), suggests just a small overlap between the eigenstates of the master modes and those belonging to the assisted ones. It is obtained from the relation 

\begin{equation}
<0\vert\left(\hat{a}_{0\omega'}\hat{a}_{0\omega''}\hat{a}^+_{\omega_k'}\hat{a}^+_{\omega_k''}\right)\vert0>=\epsilon\delta(\omega'-\omega_k'')\delta(\omega''-\omega_k').
\end{equation}
The parameter $\epsilon$ indicates the soft overlapping. When there is no overlap, then $\epsilon=0$ exactly. Besides these conditions, we have to consider $d\omega_{m}$ to be the spectrum of frequencies for which there is an overlap between $\omega$ and $\omega_k$. This is equivalent to say $d\omega_{m}=\delta(\omega-\omega_k)d\omega$, inside the interval of frequencies for which $\epsilon\neq0$. From eq. (\ref{Hal}), we can perceive that the effective gap is given by

\begin{equation}   \label{effelalala}
E_k^{eff}=E_k\left(1-\alpha\left(\frac{\hat{n}_0\hat{n}_k-\hat{n}_0^e\hat{n}_k^e-\epsilon\hat{n}_{0, kmix}^{e\beta}\hat{n}^{e\alpha}_{0, kmix}}{\hat{n}_k-\hat{n}_k^e}\right)\right).
\end{equation}
Note that the factor $\hat{n}_k-\hat{n}_k^e$ in eq. (\ref{Hal}) is just the amount of remaining particles corresponding to the assisted modes. Evidently, the evaporation of the assisted modes affects the information storage because there will be less neurons available for storing the information once this process starts. From eq. (\ref{effelalala}), we can see that the simultaneous evaporation of both, the assisted as well as the master modes, can affect the effective gap of the assisted modes. If there is no evaporation of both modes, then the perceived effective gap is still defined in eq. (\ref{gap}). Note that the effective gap depends on the net amount of assisted modes remaining in the system during the process of evaporation through the term $\hat{n}_k-\hat{n}_k^e$. It is interesting to notice that if there is no evaporation of assisted modes, then we recover the ordinary effective gap (\ref{gap}), independent on the number of master modes evaporated. This conclusion can however change if we consider higher-order terms in the Hamiltonian. On the other hand, if only assisted modes evaporate, then eq. (\ref{effelalala}) becomes

\begin{equation}   \label{effelalala2}
E_k^{eff}=E_k\left(1-\alpha\left(\frac{\hat{n}_0\hat{n}_k}{\hat{n}_k-\hat{n}_k^e}\right)\right).
\end{equation}
Note that from this expression it is clear that the evaporation of the assisted modes helps the synapse to become more gravitatory. The reason is that for the same amount of a highly occupied master mode, it is necessary to assist less modes. Then it becomes easier to make the remaining modes gapless. However, at the same time there are less modes where the information could be stored, making then it more difficult to store patterns of information. Note that when we have simultaneous evaporation of both modes, the synapse changes from gravitatory to inhibitory if the following condition is satisfied

\begin{equation}   \label{effelalala3}
\hat{n}_0\hat{n}_k<\hat{n}_0^e\hat{n}_k^e+\epsilon\hat{n}_{0, k\; mix}^{e\beta}\hat{n}^{e\alpha}_{0, k\; mix}.
\end{equation}
Note that since $\hat{n}_0^e$ and $\hat{n}_k^e$ can never be larger than $\hat{n}_0$ and $\hat{n}_k$ respectively, then the previous condition can only be satisfied in two situations. The first one, is when the black-hole is in its last stage, such that $\hat{n}_0^e\to \hat{n}_0$ and $\hat{n}_k^e\to \hat{n}_k$. The second situation is when the overlap between the master modes and the assisted ones is big enough such that $\epsilon$ becomes large enough in order to satisfy the condition (\ref{effelalala3}). In any case, we can observe that only the simultaneous evaporation of the master modes and the assisted ones might change the condition over the synapse from gravitatory to inhibitory.  

\section{Conclusions}   \label{BHBF5}          

In this letter we have demonstrated that the simultaneous evaporation of the master modes and the assisted one, under some special circumstances might change the character of the synapse. During the evaporation process the synapse might change from excitatory to inhibitory. This cannot happens with the isolated evaporation of the master modes nor with the isolated evaporation of the assisted modes. Only the combined evaporation of both modes might change the character of the synapse. When only assisted modes evaporate, the synapse becomes more gravitatory in time. This facilitates the homework of the master neuron for making the assisted modes gapless. However, at the same time, the evaporation of the assisted modes means that less patterns of information can be stored because now there are less available states after evaporation. Finally, if only master modes evaporate, the standard conditions for getting the critical condition of the system are recovered. These previous results are valid when we consider the Hamiltonian up to the second order in the occupation number. Here we consider that if the assisted modes cannot evaporate, then $\hat{n}_k^{e}=0$, as well as $\hat{n}_{0, k\; mix}^{e\beta}=0$ for the mix mode. The same number representing the evaporation of the mix modes vanishes when $\hat{n}_0^e=0$. This is the case because this mix number goes to zero if $\beta^{(0)}_{\omega, \omega'}=0$ or if $\beta^{(k)}_{\omega, \omega'}=0$ as can be observed from eqns. (\ref{security}).

\end{document}